\documentclass[preprint,preprintnumbers,amsmath,amssymb]{revtex4}
\usepackage{booktabs}
\usepackage{mathrsfs}
\usepackage{epsfig}
\usepackage{graphicx}% Include figure files
\usepackage{dcolumn}% Align table columns on decimal point
\usepackage{bm}% bold math
\usepackage{amsmath}
\usepackage{slashed}       % slashed p

\let\jnfont=\rm
\def\NPB#1,{{\jnfont Nucl.\ Phys.\ B }{\bf #1},}
\def\PLB#1,{{\jnfont Phys.\ Lett.\ B }{\bf #1},}
\def\EPJC#1,{{\jnfont Eur.\ Phys.\ Jour.\ C }{\bf #1},}
\def\PRD#1,{{\jnfont Phys.\ Rev.\ D }{\bf #1},}
\def\PRL#1,{{\jnfont Phys.\ Rev.\ Lett.\ }{\bf #1},}
\def\MPLA#1,{{\jnfont Mod.\ Phys.\ Lett.\ A }{\bf #1},}
\def\JPG#1,{{\jnfont J.\ Phys.\ G}{\bf #1},}
\def\CTP#1,{{\jnfont Commun.\ Theor.\ Phys.\ }{\bf #1},}
\def\ZPC#1,{{\jnfont Z.\ Phys.\ C }{\bf #1},}
\def\JHEP#1,{{\jnfont JHEP \ }{\bf #1},}
\def\lsim{\raise0.3ex\hbox{$<$\kern-0.75em\raise-1.1ex\hbox{$\sim$}}}
\def\gsim{\raise0.3ex\hbox{$>$\kern-0.75em\raise-1.1ex\hbox{$\sim$}}}

\begin{document}

\title{Charged Higgs bosons in the NMSSM under current LHC constraints}

\author{Zhaoxia Heng, Lin Guo, Pengqiang Sun, Wei Wei}

\affiliation{
  College of Physics and Materials Science,
        Henan Normal University, Xinxiang 453007, China
      \vspace{1cm}}

\begin{abstract}
Charged Higgs boson is a crucial prediction of new physics beyond the SM. In this work, we perform  a comprehensive scan over the parameter space of NMSSM considering various experimental constraints including the direct search limits from the 13 TeV LHC, and consider the scenario that the next-to-lightest CP-even Higgs boson is SM-like. We find that the masses of charged Higgs bosons can be as light as 350 GeV, the lightest CP-even Higgs boson $h_1$ is predominantly singlet and can be as light as 48 GeV, and the lightest CP-odd Higgs boson $a_1$ is also singlet-dominated and can be as light as 82 GeV.  The charged Higgs bosons mainly decay to $t\bar{b}$ or $\bar{t} b$, but the branching ratio of the exotic decays $H^\pm\to W^\pm h_1$ and $H^\pm\to W^\pm a_1$ can maximally reach to 20\% and 11\%, respectively, which can be used to distinguish the NMSSM from MSSM. Such a heavy charged Higgs boson is unaccessible at the 13 TeV LHC with a luminosity of 36.1 $\rm fb^{-1}$ and its detection needs higher energy and/or higher luminosity.
\end{abstract}

\maketitle

\section{Introduction}
Both the ATLAS \cite{1207ATLAS} and CMS \cite{1207CMS} collaborations at the large Hadron Collider (LHC) announced the discovery of a Higgs boson with mass about 125 GeV in 2012, which implies that the standard model (SM) of elementary particles is fully established. However, many new physics models beyond the SM with extended Higgs sectors, such as the Minimal Supersymmetric Standard Model (MSSM) \cite{MSSM}, can also accommodate a 125 GeV Higgs boson. The MSSM consists of two Higgs doublet fields, which generate the masses of up- and down-type fermions. To realize a 125 GeV Higgs, the MSSM needs large radiative corrections from the third generation squark loops \cite{MSSM-h1,MSSM-h2,MSSM-h3,MSSM-h4}, which makes the MSSM unnatural. And the MSSM also suffers from the $\mu$-problem. However, these problems can be remedied in the Next-to-MSSM (NMSSM) \cite{NMSSM}, which extends the Higgs sector with an additional Higgs singlet field $\hat S$. The effective $\mu$-term can be generated when $\hat S$ acquires vacuum expectation value (vev). The coupling between the singlet and doublet Higgs fields can easily enhance the mass of the Higgs boson to be 125 GeV without large radiative corrections \cite{h-125-1,h-125-2,h-125-3,h-125-4,h-125-5,h-125-6,h-125-7,h-125-8,
Christensen:2013dra,Kang:2012sy,Ellwanger:2011aa,King}. In contrast to the MSSM, the NMSSM has richer Higgs spectrum, which contains three CP-even Higgs bosons, two CP-odd Higgs bosons and a pair of charged Higgs  bosons $H^\pm$. Needless to say, the discovery of extra Higgs bosons along with the SM-like Higgs boson will clearly confirm the existence of new physics beyond the SM.

Because of the different interactions and decay modes from neutral Higgs bosons, the studies of charged Higgs bosons have been received more and more attentions \cite{charged higgs1,charged higgs2,charged higgs3,charged higgs4,charged higgs5}. For charged Higgs bosons lighter than top-quark, they are mainly produced through top quark decay $t\to bH^+$, and primarily decay to $\tau\nu$ and $sc$. For charged Higgs bosons heavier than top-quark, they are produced at the LHC directly through the processes $pp\to t \bar b H^\pm$, the pair production process $pp\to H^+ H^-$ and also the associated production with a neutral Higgs boson, then they may be searched via the relatively clean decay channel $H^\pm \to \tau \nu$ \cite{Aaboud:2018gjj, CMS-charged}. When kinematically allowed, the charged Higgs bosons decay to $t\bar{b}$ or $\bar{t} b$ dominantly, but it is challenging to reconstruct such events due to the large irreducible SM backgrounds \cite{Aaboud:2018cwk}. Besides the conventional search channels, the Higgs exotic decay modes \cite{Coleppa:2014hxa,exotic1,exotic2}, such as $H^\pm\to W^\pm H/A$ (H/A denotes the neutral CP-even/CP-odd Higgs boson) \cite{Coleppa:2014cca,Kling:2015uba} have studied to provide complementary detection of charged Higgs bosons. So far, the void of any charged Higgs bosons signal events limits its production and decay in a model independent way, which in turn can be used to constrain the relevant parameter space.

In this work we examine the parameter space of NMSSM considering the experimental constraints from the 125 GeV Higgs data, B-physics observables, the dark matter direct detection, and also the LHC direct search limits. We find that the charged Higgs bosons in the NMSSM can be as light as 350 GeV, and the exotic decay modes of the charged Higgs bosons are open, such as $H^\pm \to W^\pm h_1$ and $W^\pm a_1$($h_1$ and $a_1$ are the lightest CP-even and CP-odd Higgs boson, respectively), which can be used to distinguish the NMSSM from MSSM. However, the LHC with the current luminosity has not found a charged Higgs boson. To detect the charged Higgs bosons in the mass range 350-500 GeV, higher luminosity and/or higher energy collider is needed \cite{Coleppa:2014cca,Guchait:2018nkp}.

This work is organized as follows. In section II we briefly describe the NMSSM. In section III we first perform a comprehensive scan over the parameter space with the package NMSSMTools \cite{NMSSMTools1,NMSSMTools2}, then we further constrain the parameter space using the LHC direct search limits, and discuss the future detection of charged Higgs bosons at the LHC. Finally, the conclusions are drawn in section IV.

\section{Basics of the NMSSM}
As the most economic realization of supersymmetry, the MSSM consists of two Higgs doublets $\hat H_u$ and $\hat H_d$. Different from the MSSM, the NMSSM adds one extra Higgs singlet field $\hat S$. The superpotential and soft breaking terms in the Higgs sector of the NMSSM are given by
 \begin{eqnarray} \label{NMSSM-superpotential}
 W_{\rm NMSSM}&=& W_{\rm MSSM}
 + \lambda\hat{H_u} \cdot \hat{H_d} \hat{S}
 + \frac{1}{3}\kappa \hat{S^3},\\
 V_{\rm soft}^{\rm NMSSM}&=& m_{H_u}^2|H_u|^2 +  m_{H_d}^2|H_d|^2
+ m_S^2|S|^2 +(A_\lambda \lambda SH_u\cdot H_d
+\frac{A_\kappa}{3}\kappa S^3 + h.c.).
\end{eqnarray}
with $W_{\rm MSSM}$ being the superpotential of MSSM without $\mu-$term. At the tree level, the Higgs sector in the NMSSM consists of the following nine parameters:
 \begin{eqnarray}\label{parameter}
 \lambda,~\kappa,~\tan\beta,~ \mu_{\rm eff},~
 A_\lambda,~ A_\kappa,~ m_{H_u}^2,~ m_{H_d}^2,~ m_S^2.
\end{eqnarray}
where $\tan\beta=v_u/v_d$ and $\mu_{\rm eff}=\lambda v_s$ with $v_u,v_d,v_s$ denoting the vev of $H_u$, $H_d$ and $S$. The parameters $m_{H_u}^2, m_{H_d}^2, m_S^2$ can be determined by the minimization conditions of the scalar potential, so six independent parameters are left. Usually, the following six parameters are chosen as input parameters,
\begin{eqnarray}\label{parameter}
 \lambda,~ \kappa,~ \tan\beta,~ \mu_{\rm eff},~
 M^2_A=\frac{2\mu_{\rm eff}}{\sin2\beta}
  (A_\lambda+\kappa v_s),~ A_\kappa.
\end{eqnarray}

Assume $H_1 = \cos \beta H_u -  \varepsilon \sin \beta H_d^\ast$,
$H_2 = \sin \beta H_u + \varepsilon \cos \beta H_d^\ast$ with $\varepsilon$ being two-dimensional anti-symmetric matrix with off-diagonal elements of (1,-1), the Higgs fields in the NMSSM can be written as  \cite{Miller:2003ay,Kang:2012sy}:
\begin{eqnarray}
H_1 = \left ( \begin{array}{c} H^+ \\
       \frac{S_1 + i P_1}{\sqrt{2}}
        \end{array} \right),~~
H_2 & =& \left ( \begin{array}{c} G^+
            \\ v + \frac{ S_2 + i G^0}{\sqrt{2}}
            \end{array} \right),~~
H_3  = s +\frac{1}{\sqrt{2}} \left(  S_3 + i P_2 \right),
\end{eqnarray}
where $G^+$ and $G^0$ are Goldstone bosons. Obviously, the field $H_2$ corresponds to the SM Higgs field. At tree-level, the mass matrices ${\cal M}^2_S$ (under the basis ($S_1$, $S_2$, $S_3$))and ${\cal M}^2_P$ (under the basis ($P_1$, $P_2$)) are given by, respectively,
 \begin{eqnarray}
({\cal M}^2_S)_{11}&=&M_A^2+(m_Z^2-\lambda^2v^2)\sin^22\beta,\\
({\cal M}^2_S)_{12}
&=&-\frac{1}{2}(m_Z^2-\lambda^2v^2)\sin4\beta,\\
({\cal M}^2_S)_{13}
&=&-(M_A^2\sin2\beta+\frac{2\kappa\mu^2}{\lambda})\frac{\lambda v}{\mu}\cos2\beta,\\
({\cal M}^2_S)_{22}
&=&m_Z^2\cos^22\beta+\lambda^2v^2\sin^22\beta,\label{22}\\
({\cal M}^2_S)_{23}&=& 2 \lambda \mu v \left[1 - (\frac{M_A \sin 2\beta}{2 \mu} )^2
-\frac{\kappa}{2 \lambda}\sin2\beta\right],\\
({\cal M}^2_S)_{33}&=& \frac{1}{4} \lambda^2 v^2 (\frac{M_A \sin 2\beta}{\mu})^2
+ \frac{\kappa\mu}{\lambda} (A_\kappa +  \frac{4\kappa\mu}{\lambda} )
- \frac{1}{2} \lambda \kappa v^2 \sin 2 \beta, \\
 ({\cal M}^2_P)_{11} &=& M_A^2 \\
  ({\cal M}^2_P)_{12} &=& \frac{1}{2}(M_A^2\sin2\beta-6\lambda\kappa v_s^2)\frac{v}{v_s}\\
  ({\cal M}^2_P)_{22} &=& \frac{1}{4}(M_A^2\sin2\beta+6\lambda\kappa v_s^2)
  (\frac{v}{v_s})^2\sin2\beta -3\kappa v_s A_\kappa.
\end{eqnarray}
Using the rotation  matrices $U^S$ and $U^P$ to diagonalize the matrices ${\cal M}^2_S$ and ${\cal M}^2_P$, respectively, the CP-even and CP-odd Higgs mass eigenstate can be obtained by $h_i=\sum\limits_{j=1}^3 U^S_{ij}S_j$,
$a_i=\sum\limits_{j=1}^2 U^P_{ij}P_j$.
We call the scalar $h_i$ with largest $S_2$ component being the SM-like Higgs boson $h$ and take $m_{h_1}<m_{h_2}<m_{h_3}$ and $m_{a_1}<m_{a_2}$.
The element of mass matrix ${\cal M}^2_{22}$ indicates that
the mass of SM-like Higgs boson receives an additional contribution
$\lambda^2v^2\sin^22\beta$ in contrast with that of MSSM. Furthermore, the ($S_2,S_3$) mixing can also raise the mass of SM-like Higgs boson if $({\cal M}^2_S)_{22}>({\cal M}^2_S)_{33}$. This case corresponds to the next-to-lightest CP-even Higgs
boson being SM-like. In this work, we only consider the scenario with $h_2$ being SM-like Higgs boson.

The masses of charged bosons $H^\pm$ at tree-level can be obtained as \cite{charged-mass1,charged-mass2}
\begin{eqnarray}
  m^2_{H^\pm} = M^2_A+m^2_W-\lambda^2v^2  \label{mass-charged}.
\end{eqnarray}
and the couplings of charged Higgs boson with the third generation fermions are as follows,
\begin{eqnarray}
  g_{H^- t\bar{b}} &=& \frac{g_2}{\sqrt{2}m_W}(m_b\tan\beta P_L
  + m_t\cot\beta P_R) \nonumber\\
  g_{H^-\bar\tau\nu} &=& \frac{g_2}{\sqrt{2}m_W}m_\tau\tan\beta P_L \label{charged-coupling}.
\end{eqnarray}

In the neutralino sector, the gauginos $\tilde{B}$ and $\tilde{W}^0$ mix with the neutral Higgsinos $\tilde{H}^0_u$, $\tilde{H}^0_d$ and singlino $\tilde{S}$ to form a symmetric $5\times 5$ mass matrix ${\cal M}_0$. In the basis $\psi=(-i\tilde{B},-i\tilde{W}^0,\tilde{H}^0_d,\tilde{H}^0_u,\tilde{S})$
the matrix ${\cal M}_0$ is given by \cite{NMSSM}
\begin{eqnarray}
 {\cal M}_0 =
 \left(\begin{array}{ccccc}
  M_1 & 0 & -\frac{g_1 v_d}{\sqrt{2}} &
 \frac{g_1 v_u}{\sqrt{2}} & 0 \\
   & M_2 & \frac{g_2 v_d}{\sqrt{2}} &
    -\frac{g_2 v_u}{\sqrt{2}} & 0 \\
      &  & 0 & -\mu_{\rm eff} & -\lambda v_u \\
       &  &  & 0 & -\lambda v_d \\
         &  &  &  & 2\kappa v_s \\
             \end{array} \right)
\end{eqnarray}
with $M_1$ ($M_2$) being Bino (Wino) mass term, and $g_1$, $g_2$ being SM gauge couplings. Using the rotation matrix $N$ to diagonalize the matrix ${\cal M}_0$, the mass eigenstates $\tilde{\chi}_i^0$ are written as $\tilde{\chi}_i^0 =\sum\limits_{j=1}^5 N_{ij}\psi_j$, and the masses of neutralino are arranged in ascending order.

As in the MSSM, the charged gauginos $\tilde{W}^+$, $\tilde{W}^-$ mix with the charged Higgsinos $\tilde{H}^+_u$, $\tilde{H}^-_d$ to form two mass eigenstates called charginos
$\tilde{\chi}_i^{\pm} (i=1,2)$
with $m_{\tilde{\chi}_1^{\pm}} < m_{\tilde{\chi}_2^{\pm}}$.

\section{Calculations and Numerical Results}
\subsection{Scan strategies and preliminary results}
We use the package NMSSMTools \cite{NMSSMTools1,NMSSMTools2} to obtain the particle spectrum, decay branching ratios of Higgs bosons and relevant couplings. Firstly we fix the gluino mass to be 1900 GeV and the soft breaking parameters in the first two generation squark to be 2 TeV. We assume the soft breaking parameters for the left- and right-handed states in the slepton sector to be $m_{\tilde{l}}=$ 350 GeV and $A_\tau=A_e=A_\mu = $ 1500 GeV. The absence of a Landau pole below the GUT scale implies that $\lambda,\kappa\leq 0.7$. The lower limit on chargino masses from LEP is 103.5 GeV and the naturalness usually requires a low value of $\mu$, so we choose 100 GeV $\leq \mu \leq$ 500 GeV. Considering the constraints from  LHC search for electroweakinos, we choose $M_{2} \leq$ 1 TeV. In this work, we focus on searching for a light charged Higgs boson, so we require $M_A\leq 500{\rm GeV}$.
Then we perform a comprehensive scan over the following parameter regions:
\begin{eqnarray}
\begin{split}
 & 0.001 < \lambda,\kappa \leq 0.7, \quad 1.5 \leq \tan\beta \leq 60, \\
 & 100 {\rm GeV} \leq \mu \leq 500{\rm GeV},
 \quad 0 \leq M_{P} \leq 1000{\rm GeV} \\
 & 50{\rm GeV} \leq M_{1} \leq 400{\rm GeV},
 \quad 50{\rm GeV}\leq M_{2} \leq 1000{\rm GeV}, \\
 & 300{\rm GeV} \leq M_{Q_3}, M_{U_3}(M_{D_3}) \leq 2000{\rm GeV},\\
 & -4000{\rm GeV} \leq A_{t}=A_{b} \leq 4000{\rm GeV},
 \quad 100{\rm GeV} \leq M_{A} \leq 500{\rm GeV}.
\end{split}
\end{eqnarray}
where $M_P$ is the singlet diagonal element of CP-odd Higgs mass matrix.

We pick up the samples that satisfy the following constraints:
\begin{itemize}
  \item The direct mass bounds on Higgs bosons and sparticles from LEP and Tevatron experiments.
  \item The constraints from B-physics observables such as the branching ratio of processes $B_s\to \mu^+\mu^-$, $B\to X_s\gamma$ and $B_u\to \tau\nu$. We require their theoretical predications is within 2$\sigma$ range of the corresponding experimental values.
  \item The constraints from the dark matter relic density with $0.1068<\Omega h^2<0.13057$ consistent with the Planck measurement \cite{dm}. Because of the existence of blind spots
      \cite{blind spot1,blind spot2} for neutralino dark matter in the NMSSM, we do not consider the constraints from dark matter direct detection experiments, such as LUX and XENON-1T experiments.
  \item  The constraints from Higgs data. We assume the next-to-lightest CP-even Higgs boson being SM-like and satisfying 122 GeV $\leq m_{h_2}\leq$ 128 GeV. We consider the constraints from LEP, Tevatron and LHC on the direct searches for neutral and charged Higgs bosons using the package \texttt{HiggsBounds} \cite{HiggsBounds1,HiggsBounds2} and perform the fit for the 125GeV Higgs data using the package \texttt{HiggsSignals} \cite{HiggsSignals1,HiggsSignals2,HiggsSignals3}.
\end{itemize}

\begin{figure}[htbp]
\includegraphics[width=12cm]{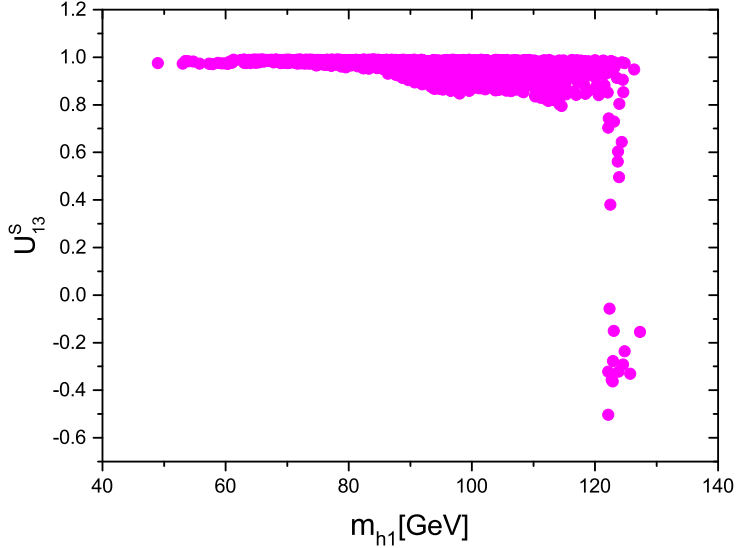}
\caption{Surviving samples in the $U^S_{13}-m_{h_1}$ plane.}
\label{fig1}
\end{figure}

\begin{figure}[htbp]
\includegraphics[width=12cm]{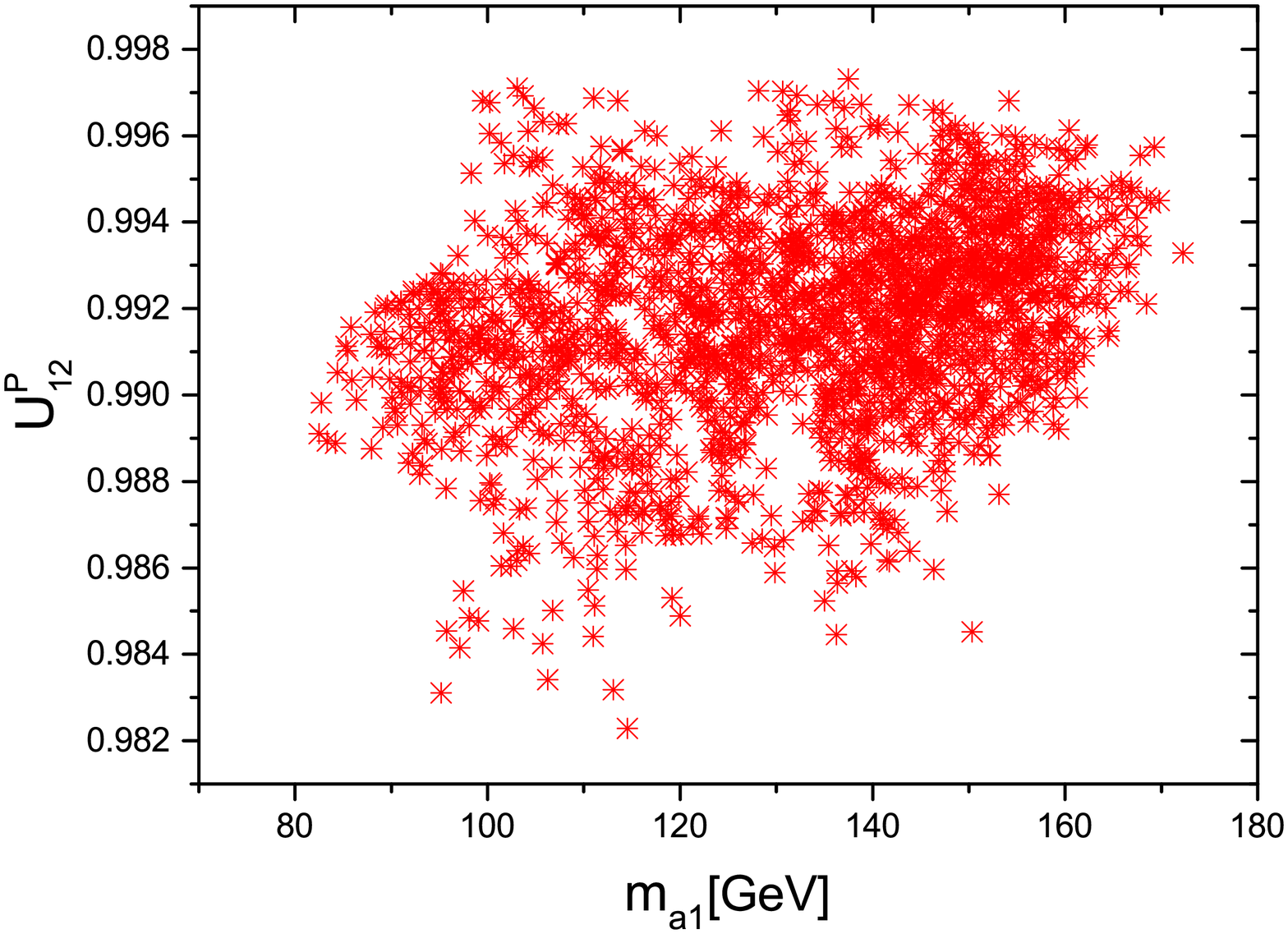}
\caption{Surviving samples in the $U^P_{12}-m_{a_1}$ plane.}
\label{fig2}
\end{figure}

\begin{figure}[htbp]
\includegraphics[width=12cm]{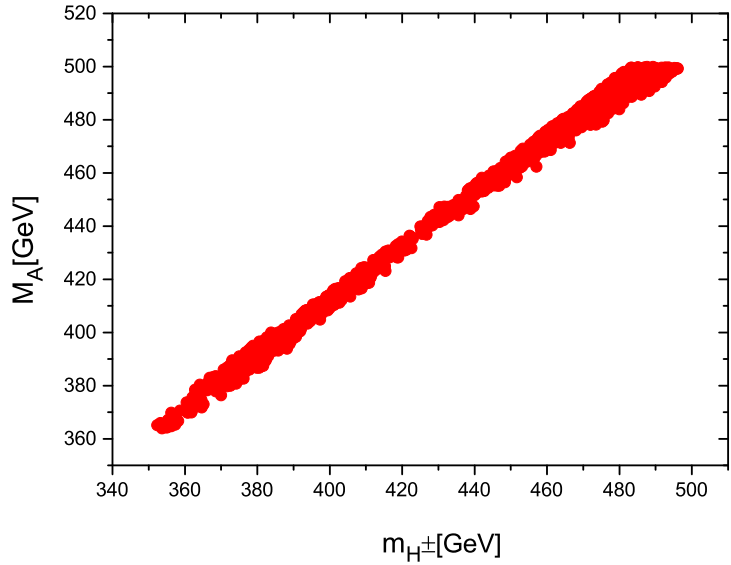}
\caption{Surviving samples in the $M_A-m_{H^\pm}$ plane.}
\label{fig3}
\end{figure}

In Fig.\ref{fig1} and Fig.\ref{fig2} we display the surviving samples in the $U^S_{13}-m_{h_1}$, $U^P_{12}-m_{a_1}$ planes, respectively.
Fig.\ref{fig1} shows that $h_1$ is predominantly singlet and can be as light as 48 GeV. Fig.\ref{fig2} shows that $a_1$ is also dominated by singlet component and can be as light as 82 GeV.
In Fig.\ref{fig3} we display the surviving samples in the $M_A-m_{H^\pm}$ plane. The figure shows that the masses of charged Higgs bosons can be as light as 350 GeV and are highly correlated with $M_A$, which can be seen clearly from Eq.(\ref{mass-charged}). The mass upper bound of charged Higgs bosons is due to the requirement of $M_A\leq 500{\rm GeV}$.

\begin{figure}[htbp]
\includegraphics[width=8cm]{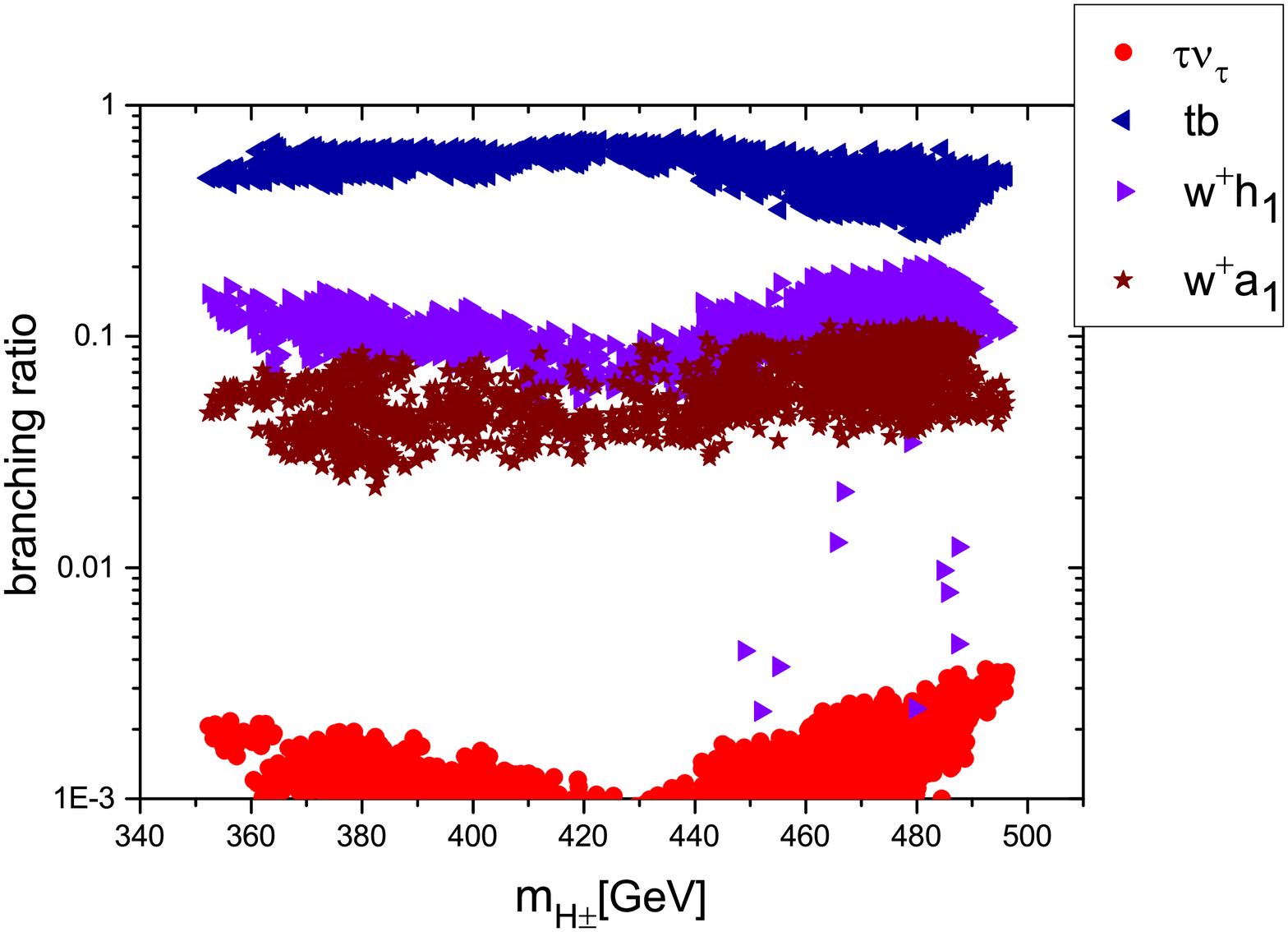}
\includegraphics[width=8cm]{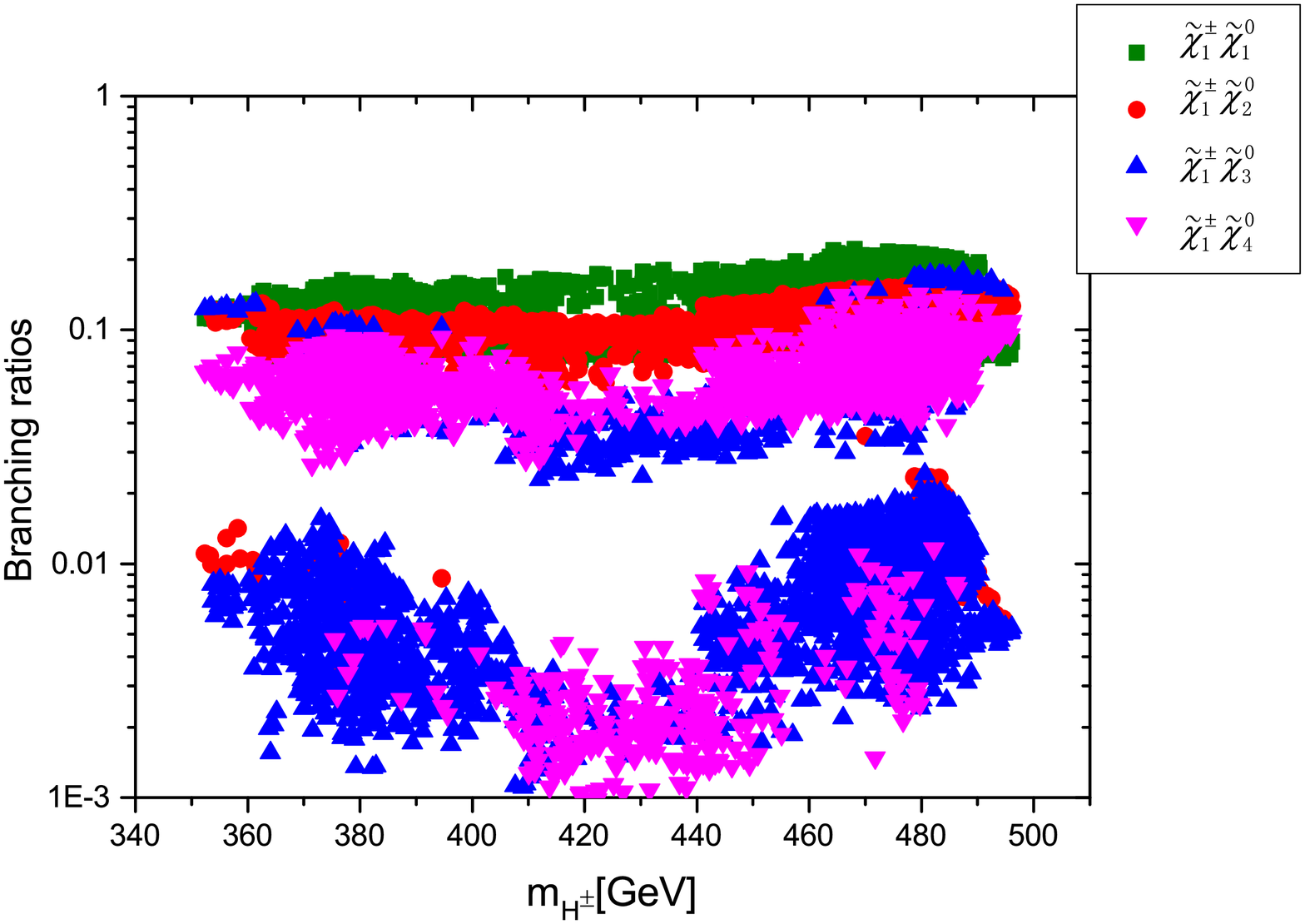}
\vspace{-0.2cm}
\caption{Branching ratios of different decay channels of charged Higgs bosons versus $m_{H^\pm}$.}
\label{fig4}
\end{figure}

In Fig.\ref{fig4} we show the different decay channels of charged Higgs bosons in the NMSSM. From the figure, we can see that the dominant decay channel is $H^+\to t\bar{b}$, which varies from 71\% to 28\%. Although the decay channel $H^\pm\to \tau^\pm \nu_\tau$ may be dominant for light charged Higgs bosons, in our case it is rather small. Firstly it is because the $H^\pm \tau\nu_\tau$ coupling is suppressed for small $\tan\beta$, which can be seen from Eq.(\ref{charged-coupling}). Secondly, some exotic decay channels are open, such as $H^\pm\to W^\pm h_1$, $H^\pm \to W^\pm a_1$ and
$H^\pm\to \tilde{\chi}^\pm_1\tilde{\chi}^0_i$.
The figure shows that $Br(H^\pm\to W^\pm h_1)$ and $Br(H^\pm\to W^\pm a_1)$ can maximally reach to 20\% and 11\%, respectively, and the branching ratio of
$H^\pm\to \tilde{\chi}^\pm_1\tilde{\chi}^0_1$ can reach to about 22\%. As is well known, the branching ratio of the decay $H^\pm\to W^\pm h_1/a_1$ is strongly dependent on the mass of $h_1/a_1$ and the coupling $W^\pm H^\mp h_1/a_1$, which is directly proportional to the doublet component of $h_1/a_1$. Although $h_1$ and $a_1$ are dominated by singlet component, they can be very light (see Fig.\ref{fig1} and Fig.\ref{fig2}), therefore the branching ratio of $H^\pm\to W^\pm h_1/a_1$ can be sizeable. In the MSSM, the relationship $m^2_{H^\pm}=M^2_A+m^2_W$ ensures that the decay $H^\pm\to W^\pm A$ is strongly suppressed in most of the parameter space. Therefore, the detection of the decay channel $H^\pm\to W^\pm h_1/a_1$ can be used to distinguish the NMSSM from MSSM.

\subsection{The LHC direct search limits}
Both the ATLAS and CMS collaborations have searched for the charged Higgs bosons via the production process $t\to bH^+$ or $pp\to tbH^\pm$, subsequently charged Higgs bosons decay to $\tau\nu$ \cite{Aaboud:2018gjj, CMS-charged}. With $\mathcal{L}=36.1 \rm fb^{-1}$ at 13 TeV, the observed 95\% C.L. upper limits from the ATLAS \cite{Aaboud:2018gjj} on the production cross section of $H^\pm$ times $Br(H^\pm\to \tau^\pm\nu)$ is between 4.2 pb and 2.5 fb for $\rm 90 GeV\lesssim m_{ H^\pm }\lesssim 2000 GeV$. Interpreting in the hMSSM scenario of the MSSM, charged Higgs bosons lighter than 160 GeV are excluded for all $\tan\beta$ values, and charged Higgs bosons lighter than 1100 GeV are excluded when $\tan\beta$ = 60. The CMS gave the similar results \cite{CMS-charged}. The ATLAS also searched for charged Higgs bosons via the process  $pp\to tbH^\pm$ and then decaying to $tb$ \cite{Aaboud:2018cwk}. With $\mathcal{L}=36.1 \rm fb^{-1}$ at 13 TeV, the observed 95\% C.L. upper limits on $\sigma(pp\to tbH^\pm)\times Br(H^\pm\to tb)$ is in the range 2.9-0.07pb for $\rm 200 GeV\lesssim m_{ H^\pm }\lesssim 2000GeV$. In the hMSSM scenario, $0.5<\tan\beta<1.95$ is excluded for $200 \rm GeV \lesssim m_{ H^\pm }\lesssim 965 GeV$.

\begin{figure}[htbp]
\includegraphics[width=12cm]{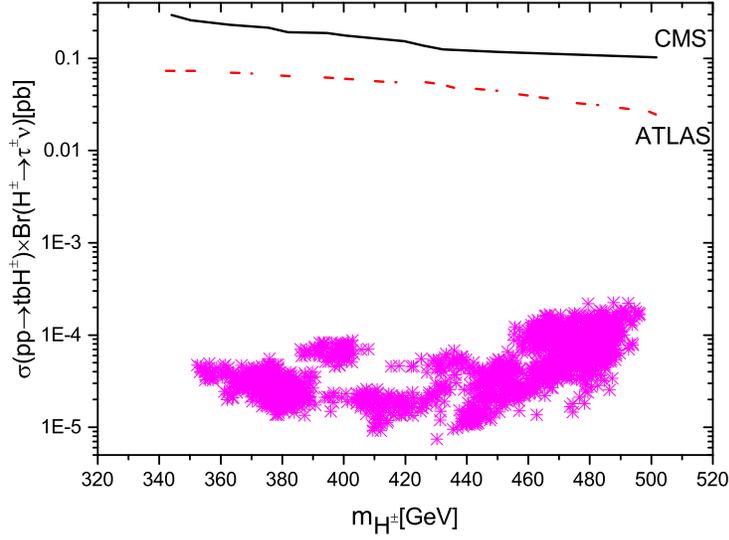}
\caption{Surviving samples in the $\sigma(pp\to tbH^\pm)\times Br(H^\pm\to\tau^\pm\nu)- m_{H^\pm}$ plane. The observed 95\% C.L. exclusion limits on $\sigma\times Br$ from ATLAS and CMS with integrated luminosity of 36.1 $\rm fb^{-1}$ at $\sqrt{s}$ = 13 TeV are also shown in the figure.}
\label{fig6}
\end{figure}

\begin{figure}[htbp]
\includegraphics[width=12cm]{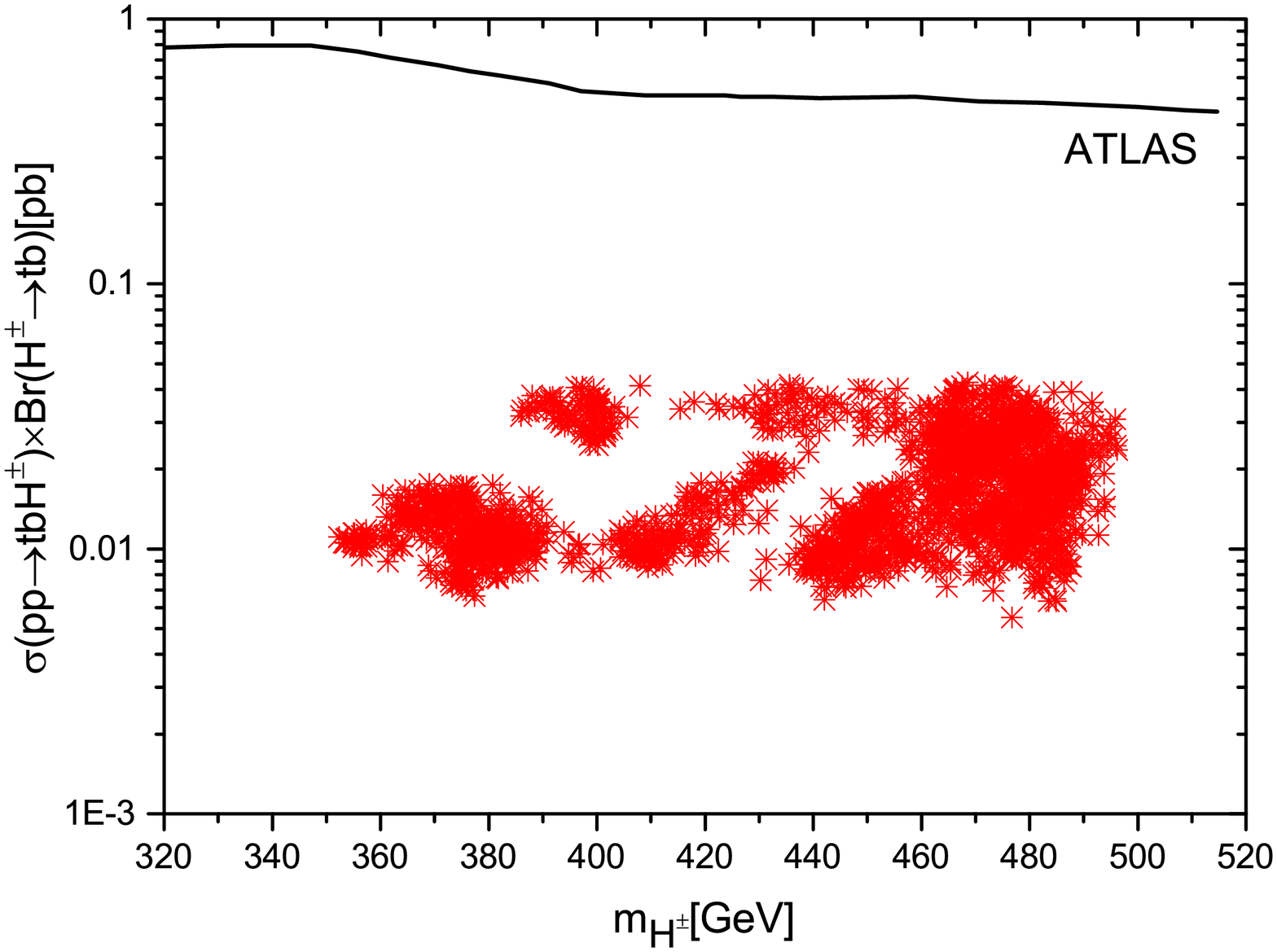}
\caption{Surviving samples in the $\sigma(pp\to tbH^\pm)\times Br(H^\pm\to tb)- m_{H^\pm}$ plane (right). The observed 95\% C.L. exclusion limits on $\sigma\times Br$ from ATLAS with integrated luminosity of 36.1 $\rm fb^{-1}$ at $\sqrt{s}$ = 13 TeV are also shown in the figure.}
\label{fig7}
\end{figure}

To compare with the LHC direct search limits, we calculate the production cross section of the process $pp\to tbH^\pm$ times $Br(H^\pm\to \tau^\pm\nu)$ or $Br(H^\pm\to tb)$ at $\sqrt{s}$ =13 TeV, and show the results in Fig.\ref{fig6} and Fig.\ref{fig7}, which also shows the observed 95\% C.L. exclusion limits on  $\sigma\times Br$ from ATLAS \cite{Aaboud:2018gjj, Aaboud:2018cwk} and CMS \cite{CMS-charged} with integrated luminosity of 36.1 $\rm fb^{-1}$ at $\sqrt{s}$ = 13 TeV. The figure indicates that the $\sigma\times Br$ of the surviving samples are less than the observed 95\% C.L. exclusion limits from the ATLAS and CMS.

Due to the large irreducible SM backgrounds, we also consider the LHC direct search limits for the charged Higgs bosons produced through the electroweak processes \cite{Wang:2018hnw}:
\begin{eqnarray*}
  pp&\to &W^{\pm *}\to H^{\pm}h_i (i=1,2,3) \\
  pp&\to &W^{\pm *}\to H^{\pm}a_j (j=1,2)\\
   pp&\to &Z^*/\gamma^* \to H^+H^- .
\end{eqnarray*}
and consider the decay channel of charged Higgs bosons to
$\tau^\pm \nu$ or $W^\pm h_1/a_1$. For the heavier neutral Higgs bosons $h_3$ and $a_2$, their dominant decay channels are $h_3\to a_1Z$, $a_2\to h_1Z$. For the lighter neutral Higgs bosons $h_1$ and $a_1$, they mainly decay to $b\bar b$ with branching ratio close to 90\%, and subsequently decay to $\tau^+\tau^-$ with branching ratio close to 10\%. In order to reduce the irreducible background, we consider the decay mode of $h_1/a_1\to\tau^+\tau^-$. The cross sections of processes
$pp\to W^{\pm *}\to H^{\pm}h_1/a_1$ are strongly correlated with the coupling $W^\pm H^\mp h_1/a_1$, which is directly proportional to the doublet component of $h_1/a_1$. Since $h_1/a_1$ are singlet dominated, the cross sections of processes $pp\to W^{\pm *}\to H^{\pm}h_1/a_1$ are suppressed. So we consider the following final states,
\begin{eqnarray*}
& pp\to W^{\pm *}\to H^{\pm}h_3/a_2 \to 3\tau +\nu_\tau +Z
  ~{\rm or }~ 4\tau + Z + W^\pm, \\
& pp\to Z^*/\gamma^* \to H^+H^-\to 2\tau +\nu_\tau
   ~{\rm or }~ 4\tau + W^ + W^- \\
&   ~{\rm or }~ 3\tau +\nu_\tau + W^+/W^-.
\end{eqnarray*}
We generate the parton level signal events using \texttt{MG5\_aMC-2.4.3} \cite{madgraph} with \texttt{PYTHIA6}~\cite{Torrielli:2010aw} performing parton showering and hadronization, then we use \texttt{CheckMATE-2.0.7} \cite{Dercks:2016npn} with all the analysis at 13 TeV LHC to perform simulations. For most of the surviving samples, $R<1$, where $R=\max\limits_i {\{R_i\}}$ with $R_i$ being the limit for each searching analysis by ATLAS or CMS. $R_i=\max\limits_j \{R_{i,j}\}$ with $j$ standing for each signal region at one analysis, and $R_{i,j}=\frac{S}{S^{95}_{obs}}$ with $S$ being predicted events number of the model and $S^{95}_{obs}$ being upper limit of events number at 95\% confidence level.

Combined the LHC direct search limits discussed above, the charged Higgs bosons can be as light as 350 GeV in the NMSSM.

\subsection{The future detection of charged Higgs bosons}
When charged Higgs bosons heavier than top quark, it dominantly decay to $t\bar{b}$ or $\bar{t} b$. But such signal events are overwhelmed by the large SM background. And the branching ratio of the clean decay channel $ H^\pm\to \tau^\pm\nu$ is rather small. Therefore, it is challenging to search for charged Higgs bosons through the conventional search channels. Ref.\cite{Coleppa:2014cca} studied the exotic decay channel of charged Higgs bosons $H^\pm\to AW^\pm$ in two Higgs doublet model (2HDM). For $m_A$ = 70 GeV with small and large $\tan\beta$, a charged Higgs boson lighter than 400 GeV may be discovered with $\mathcal{L}=300 \rm fb^{-1}$ at the LHC. In the $m_{H^\pm}-\tan\beta$ plane, the charged Higgs bosons with masses extended to 600 GeV can be excluded at the 95\% C.L.. Adopting the multivariate analysis technique to improve the signal sensitivity, Ref.\cite{Guchait:2018nkp} studied the signature of a heavier charged Higgs boson for the decay mode $H^+\to t\bar{b}$ with both the hadronic and leptonic final states. For $\tan\beta$ =3, the charged Higgs bosons in 2HDM with mass between 300 and 600 GeV may be observable at the LHC with $\mathcal{L}=1000 \rm fb^{-1}$. Therefore,
the future higher luminosity and/or higher energy colliders would be useful for the detection of charged Higgs bosons \cite{Aboubrahim:2018tpf, Kling:2018xud, Cepeda:2019klc}.

\section{Conclusion}
Due to the different peculiarity from the neutral Higgs bosons, the charged Higgs bosons have been received more and more attentions. The search for charged Higgs bosons would be a crucial signal of new physics beyond the SM. We perform  a comprehensive scan over the parameter space of the NMSSM considering various experimental constraints including the direct search limits from the 13 TeV LHC, and pick up the samples with the next-to-lightest CP-even Higgs boson being SM-like. We find that the masses of charged Higgs bosons are larger than 350 GeV, the lightest CP-even Higgs boson $h_1$ is predominantly singlet and can be as light as 48 GeV, and the lightest CP-odd Higgs boson $a_1$ is also singlet-dominated and can be as light as 82 GeV. We also discuss the different decay channels of the charged Higgs bosons. The charged Higgs bosons mainly decay to $t\bar{b}$ or $\bar{t} b$, but some exotic decay channels are open. The branching ratio of the decay $H^\pm\to W^\pm h_1$ and $H^\pm\to W^\pm a_1$ can maximally reach to 20\% and 11\%, respectively, which can be used to distinguish the NMSSM from MSSM.

Because of the large SM backgrounds for the decay channel $H^+\to t\bar{b}$ and rather small decay branching ratio of the decay channel
$H^\pm \to \tau^\pm\nu$, no evidence of a charged Higgs boson is found at the LHC with $\mathcal{L}=36.1 \rm fb^{-1}$ integrated luminosity. The future higher luminosity and/or higher energy colliders would be useful for the detection of charged Higgs bosons.

\section*{Acknowledgement}
We thank Prof.Junjie Cao, Dr.Liangliang Shang and Dr.Yang Zhang for helpful discussions. This work was supported in part by the National Natural Science Foundation of China (NNSFC) under grant No.11705048, and the Program for Innovative Research Team in University of Henan Province under grant No.19IRTSTHN018.
%%%%%%%%%%%%%%%%%%%%%%%%%%%%%%%%%%%%%%%%%%%%%%%%%%%%%%%%%%%%%%%%%%%%%%%%%
%%%%%%%%%%%%%%%%%%%%%%%%%%%%%%%%%%%%%%%%%%%%%%%%%%%%%%%%%%%%%%%%%%%%%%%%%

\end{document}